\documentstyle[aps,preprint]{revtex}
\begin{document}
\title{Logarithmic density relaxation in compaction of granular materials}
\author{Gongwen Peng\thanks{e-mail: peng@phys.ocha.ac.jp}
and Takao Ohta\thanks{e-mail: ohta@phys.ocha.ac.jp} }
\address{Department of Physics, Ochanomizu University, Tokyo 112, Japan}
%\date{today}
\maketitle
\begin{abstract}
\indent On the basis of physical considerations we propose a one--dimensional discrete lattice model for the density relaxation of granular materials under tapping. Solving the difference equation numerically, we find a logarithmic time--dependence of the density relaxation. This is in agreement with experimental results of Knight $et$ $al.$ [Phys. Rev. E {\bf 51}, 3957(1995)]. The origin of this anomalous relaxation is elucidated analytically by solving the equation of its continuum version asymptotically in time.

%Basing on physical considerations we propose a partial differential equation f%or the density relaxation of granular materials under tapping.  We solve the e%quation both numerically and analytically in the asymptotic limit. The results% give a rigorous confirmation to the logarithmic functional form obtained from% experimental data by Knight $et$ $al.$ [Phys. Rev. E {\bf 51}, 3957(1995)].
\vspace{2cm}

\noindent
PACS Numbers: 46.10.+z, 05.40.+j, 05.60.+w
\indent
\pacs{46.10.+z, 05.40.+j, 05.60.+w}

\end{abstract}

\vspace*{0.5cm}

\section{Introduction}

The slow relaxation of density in granular materials under tapping  has attracted much theoretical interest recently \cite{hans1,hans2,hans3,hans4,linz} since  Knight $et$ $al.$ \cite{Knight} carried out systematical experimental measurements of density as a function of time for a vibrated granular material. The experimental data of Knight $et$ $al.$ \cite{Knight} can be most satisfactorily fitted using a functional form:
\begin{equation}
\label{log}
\rho(t) = \rho_{\infty} - \frac{\delta \rho_{\infty}}{1 + G \ln [1 + \frac{t}{\tau}]}
\end{equation}
where $\rho(t)$ is the average density in the system at time instant $t$. Here $\rho_{\infty}$, $\delta \rho_{\infty}$, $G$ and $\tau$ are constant parameters. This four--parameter fit was obtained from experimental data without theoretical motivation. Previous theoretical models predicted different functional forms. In the model of Barker and Mehta \cite{Mehta}, particles can relax both independently, as individual particles,  and collectively, as clusters. Their model leads to a sum of two exponential terms, which is not in accord with the logarithmic form of Eq.(1). The model proposed by Hong $et$ $al.$ \cite{Hong} is based on a diffusing void picture. It predicts a power--law dependence of height reduction as a function of time, $\delta h \sim t^z$ with $z = 1$, which means that the density increases linearly until saturation.

Using frustrated Ising models, Herrmann and his collaborators \cite{hans1,hans2,hans3,hans4} gave a numerical confirmation of the logarithmic density relaxation as Eq. (\ref{log}) in their Monte--Carlo simulations. In their models, frustration plays a crucial role. They claim that frustration is generated in granular materials by the steric constraints imposed by the hard--core repulsion of neighboring grains and the subsequent interlocking. The link of the frustrated lattice gas models  
to  real granular packing is explained in Refs. \cite{hans1,hans2,hans3,hans4}. 

In this paper, we propose a one--dimensional discrete lattice model for  granular compaction under tapping based on a simple physical picture. Numerical solutions to the model equation indicate that the density relaxes in a  logarithmic manner, precisely as Eq. (1). We also solve the continuum version of the model  analytically in the asympototic limit ($t \rightarrow \infty$) to confirm Eq. (1). 

This paper is organized as follows. Section II presents the model equation. Numerical and analytical solutions to the equation are given in Sections III and IV respectively. Section V is devoted to discussion.

\section{Model Equation for Density Relaxation}

The density of particles $\rho({\bf r}, t)$ should satisfy the continuity equation:
\begin{equation}
\label{con}
\frac{\partial}{\partial t} \rho({\bf r}, t) + \nabla \cdot {\bf J} = 0
\end{equation}
where ${\bf J}$ is the current density. We may rule out any simple density--gradient term in the current density which leads to isotropic diffusion, since the thermal energy is too small to triger the motion of grains. The current density may be written as a product of density and velocity:
\begin{equation}
\label{flux}
{\bf J} = \rho({\bf r}, t) {\bf v}
\end{equation}
To specify the the functional form of velocity, let us consider a one--dimensional lattice model along z--direction. Gravity is along the negative z--axis. In the absence of tapping, motion of grains under the action of gravity is possible only when the geometric constraint is satisfied, i.e., enough void space below the grain. If $\rho(z, t)$ is the particle density, the void density at the site just below is $ 1 - \rho(z-1, t)$. Therefore, the velocity is nonzero only when the ratio $\alpha = \frac{\rho(z, t)}{1 - \rho(z-1, t)}$ less than $1$. This means a step function for the velocity $v$:
\begin{equation}
v = \left\{ \begin{array} {lcl} \displaystyle - v_0&,&\alpha \le 1\\
                          \displaystyle 0  &,&\alpha > 1 \end{array}\right.
\end{equation}
in the absence of tapping. The effect of tapping in granular compaction is to overcome the bottleneck effect, i.e., to make the geometric constraints not as strict as the above. Motion of grains is also possible when $ \alpha > 1 $ with the help of tapping. We may therefore replace the above step function with a peak function of an exponential form:
\begin{equation}
\label{vel}
v = - D e^{-\alpha/\gamma} 
\end{equation}
where $\gamma$ and $D$ are constants dependent of the vibrational intensity $\Gamma = A\omega^2/g$ with the amplitude $A$ and the frequency $\omega$ of oscillation and $g$ the gravitational constant. The precise $\Gamma$--dependences of $D$ and $\gamma$ are beyond the scope of the present theory. Substituting Eq. (\ref{vel}) and Eq. (\ref{flux}) into Eq. (\ref{con}), we obtain our model equation for density relaxation:
\begin{equation}
\label{model}
\frac{\partial}{\partial t} \rho(z, t) = D \frac{\partial}{\partial z} F(z,t)
\end{equation}
with
\begin{equation}
\label{current}
F(z,t) = 
\rho(z, t) \exp{(-\frac{\rho(z, t)}{\gamma(1 - \rho(z-\xi, t))})}
\end{equation}
where $\xi$ is infinitesimal in the continuum limit. This model is defined more definitely on a lattice:
\begin{equation}
\label{latticemodel}
\frac{\partial \rho(z, t)}{\partial t} = D ( F(z+1, t) - F(z, t) )
\end{equation}
where $F(z, t)$ is given by Eq. (\ref{current}) putting $\xi = 1$.
%in the discrete lattice model. Its continuum version is
%\begin{equation}
%\label{current2}
%F(z,t) = 
%\rho(z, t) \exp{(-\frac{\rho(z, t)}{\gamma(1 - \rho(z, t) - \xi 
%\frac{\partial \rho}{\partial z}(z^{-}, t))})}
%\end{equation}
%where $\frac{\partial \rho}{\partial z}(z^{-}, t)$ means derivative from the l%eft side and $\xi$ is a constant parameter.

\section{Numerical Solutions}

In order to solve numerically Eq. (\ref{model}), we supply two fixed boundary conditions, $\rho(0, t) = 1$ at the bottom and $\rho(L+1, t) = 0$ at the top of the system. The following discrete version of Eq. (\ref{latticemodel}) ensures no flux at the boundaries, i.e., conservation of particle density as a whole is ensured at each time step
\begin{equation}
\label{rule}
\rho(z, t+1) = \rho(z, t) + D ( F(z+1, t) - F(z, t) )
\end{equation}
with $F(z, t)$ given by Eq. ($\ref{current}$) with $\xi = 1$. 

We update \{$\rho(z, t), z=1, 2, \cdots, L$\} according to Eq. (\ref{rule}) from random initial conditions with \{$\rho(z, 0), z=1, 2, \cdots, L$\} equal to random numbers distributed uniformly from $\rho_{l}$ to $\rho_{u}$. Fig. 1 shows snapshots of density profiles at three different time steps. We see that as particles move downward voids move upward and pile at the top. As soon as we start updating from a random initial configuration, a sharp interface  between a particle phase and a void phase begins to emerge. Its position can be easily identified as the location where density changes  sharply from non--zero value to zero. Since  the position of the interface takes only integer
 values on the lattice and  information extracted from it is therefore limited, we resort to calculate the position of center of mass
\begin{eqnarray}
H(t) & = &\int_{1}^{L} z \rho(z, t) dz \nonumber \\[3mm]
     & = &\sum_{z=1}^{L} z \rho(z, t).
\end{eqnarray}
We take it for granted that the average density measured in experiments is  proportional to the inverse of $H(t)$. We determine the proportional prefactor as follows. The minimum value of H can be obtained from a density distribution of   a step function, i.e., 
\begin{equation}
\label{step}
\rho(z, t) = \left\{ \begin{array} {lcl} \displaystyle 1&,& z \le B\\
                          \displaystyle 0  &,& z > B \end{array}\right.
\end{equation}
which is actually a static  solution to the model equation Eq. (\ref{rule}). Here
 $B$ is determined by the density conservation and is a time--independent quantity:  $B = \int_1^L \rho(z, t) dz$. The value of $H$ corresponding to Eq. (\ref{step}) is $\frac{1}{2} B^2$. We therefore use the following expression for the average density
\begin{equation}
\rho(t) = \frac{B^2}{2H(t)}
\end{equation}

Starting from random initial configurations, we discard  the data for initial transient time period and do not make computation of $\rho(t)$ until at the interface between the particle phase and the void phase the density jumps from zero to some reasonably finite value.  Our density of unity corresponds to the densest close packing and it sets the scale for density. For the case of sphearical particles in reality, the densest close packing has a density of about 0.64  while the mechanically least stable configurations have the most loose value of about 0.55 \cite{dense} (in our scale 0.859).  Fig. 2 shows one typical plot of $\rho(t)$ calculated from the lattice model. The time step when we start calculating $\rho(t)$ in Fig. 2 is set to be the time origin. It is the instant when the density gap at the interface is about $0.8$ in our density scale. The numerical data in Fig. 2 can be very well fitted by the logarithmic form of Eq. (\ref{log}), which is also shown in the figure.

Fig. 3 displays two density profiles at different time steps for the particle phase while zero density in the void phase is not plotted. We find that the density profile can be fitted by the following expression
\begin{equation}
\label{z-dep}
\rho_p(z, t) = \rho_1(t) - s(t) [\log(z)]^{\beta(t)}.
\end{equation}
where $\rho_1(t)$, $s(t)$ and $\beta(t)$ are time--dependent parameters.
The fit  to the numerical data is also plotted in Fig.  3. In the whole range the  density is step-function-like
\begin{equation}
\label{z-step}
\rho(z, t) = \left\{ \begin{array} {lcl} \displaystyle\rho_p(z, t)
                  &,& z \le b(t)\\
                          \displaystyle 0  &,& z > b(t) \end{array}\right.
\end{equation}
where $b(t)$ is the location of the interface.

Eq. (\ref{z-step}) is different from the generalized Fermi--Dirac distribution  proposed by Hayakawa and Hong \cite{Hayakawa} and found  by Herrmann and his collaborators in simulations of frustrated Ising models \cite{hans3,hans4}. The generalized Fermi--Dirac distribution does not have a singularity in the density profile but changes continuously from larger values to zero rapidly within a boundary layer  whose width is determined by one of the parameters in the distribution. On the other hand, our model gives a sharp interface between the particle phase and the void phase and therefore the model does not permit any density value smaller than the most loose density for granular compaction. In this sense, we may say that in addition to the granular solid phase and the  void phase the generalized Fermi--Dirac distribution allows  a granular $gas$ phase where  density changes drastically \cite{Hayakawa,hans3,hans4}. Our model describes only the granular compaction under tapping where the grains do not fly as in a gas (and therefore no inertia terms appear in the velocity expression Eq. (\ref{vel})). 

\section{Analytical Solution in the asymptotic limit}

As we observed in the simulations, there is always a sharp interface between the particle phase and the void phase. We may therefore generally postulate the solution to Eq. (\ref{model}) as 
\begin{equation}
\label{0}
\rho(z, t) = a(z, t) \Theta(b(t) - z)
\end{equation}
where $\Theta(x)$ is the step function such that $\Theta(x)=0$ for $x \le 0$ and $\Theta(x)=1$ for $x > 0$. We now determine the time dependence of the location of the interface $b(t)$ by solving the equation in its continuum version, Eq. (\ref{model}). Substituting Eq. (\ref{0}) into the left hand side of Eq. (\ref{model}), we obtain 
\begin{equation}
\label{1}
\frac{\partial \rho}{\partial t} = \dot{a}(z,t)\Theta
   (b(t) - z) + a(z,t) \dot{b}(t) \delta (b(t) - z)
\end{equation}
where dot stands for time deriative.
The $\Theta$ function in Eq. (\ref{0}) gives us the following $F(z,t)$ in Eq. (\ref{current})
\begin{eqnarray}
F(z,t) & = & a(z,t) \Theta (b(t) - z) e^{-a(z,t)/(\gamma(1-a_{-}(z,t)))} 
\\[3mm]
    & = & f(a(z,t), a_{-}(z,t)) \Theta(b(t) - z)
\end{eqnarray}
where $a_{-}(z,t)$ stands for the value of $a$ just below $z$, i.e. $a(z-\xi, t)$ with infinitesimal $\xi$. Here 
\begin{equation}
f(a(z,t), a_{-}(z,t)) \equiv  a(z,t) e^{-a(z,t)/(\gamma(1-a_{-}(z,t)))}
\end{equation}
Thus, the right hand side of   Eq. (\ref{model}) reads
\begin{equation}
\label{2}
D \frac{\partial F(z,t)}{\partial z} = D \frac{\partial f}{\partial z} \Theta(b(t) - z) - D f \delta(b(t) - z) 
\end{equation}
Integrating the right hand sides of Eq. (\ref{1}) and  of Eq. (\ref{2}) from $z = b(t) - \Delta$ to 
$z = b(t) + \Delta$ and taking the limit of $\Delta \rightarrow 0$, we obtain
\begin{equation}
\label{atinterface}
a(z,t) \dot{b}(t) |_{z=b(t)} = -D a(z,t) e^{-a(z,t)/(\gamma(1-a_{-}(z,t)))} |_{z=b(t)}
\end{equation}

Now we make an approximation. From the simulations we know that  $a(z,t)$ depends on $z$ very weakly. The parameter $\beta(t)$ in Eq. (\ref{z-dep}) decreases as time increases. When $t \rightarrow \infty$, $\beta \rightarrow 0$. We therefore make the following aussmption in the asymptotic limit of $t \rightarrow \infty$
\begin{equation}
\label{ass1}
a(z,t) = a(t).
\end{equation}
The conservation of total density  is then expressed as
\begin{equation}
\label{ass2}
a(t) b(t) = B
\end{equation}
with the same $B$ as in Eq. (\ref{step}).

Using Eq. (\ref{atinterface}), Eq. (\ref{ass1}) and Eq. (\ref{ass2}) we have 
\begin{equation}
\label{a}
B \dot{a} = D a^2 e^{-\frac{a}{\gamma(1-a)}}
\end{equation}
Letting $a = 1 - \epsilon$ leads in the asymptotoc limit ($\epsilon \rightarrow 0$) 
\begin{equation}
\label{eps2}
\dot{\epsilon} = -c e^{-\frac{1}{\gamma \epsilon}}
\end{equation}
where $c = \frac{D e^{1/\gamma}}{B}$ is constant. Putting $x = 1/\epsilon$, we have
\begin{equation}
\label{x}
\dot{x} = c x^2 e^{-x/\gamma}.
\end{equation}

Now we integrate Eq. (\ref{x}) for $t$ from $0$ to $t$ and for $x$ from $\frac{1
}{1 - \rho_0}$ to $\frac{1}{1 - \rho}$ where $\rho_0$ and $\rho$ are the average
 density (i.e., a(t)) of the particle phase at time $0$ and $t$ respectively
\begin{equation}
\label{int_1}
\int_{\frac{1}{1-\rho_0}}^{\frac{1}{1-\rho}} x^{-2} e^{x/\gamma} dx = c \int_{0}
^{t} dt.
\end{equation}
The asymptotic solution of Eq. (\ref{int_1}) is readily obtained by using the formular
\begin{equation}
\label{ei}
Ei(-x) = -\int_{x}^{\infty} \frac{e^{-y}}{y} dy
\end{equation}
and the expansion for $x >> 1$
\begin{equation}
\label{ei2}
Ei(-x) = e^{-x}[-\frac{1}{x} + \frac{1}{x^2} + O(\frac{1}{x^3})].
\end{equation}
The final form for Eq. (\ref{int_1}) is given by
\begin{equation}
\label{d}
\gamma(1-\rho)^2 e^{1/(\gamma(1-\rho))}
 = ct +d
\end{equation} 
with  $d = \gamma(1-\rho_0)^2 e^{1/(\gamma(1-\rho_0))}$.
This can be rewritten as
\begin{equation}
\label{dlog}
\ln \gamma + 2 \ln (1- \rho) + \frac{1}{\gamma(1-\rho)} = \ln (ct + d)
\end{equation}
where, however, the first two terms are negligible compared with the third one for the left side, so that we obtain the final form of the average density 
\begin{equation}
\label{final}
\rho(t) = 1 - \displaystyle
\frac{\displaystyle\frac{1}{\gamma \ln(d)}}{\displaystyle 1 + \frac{1}{\ln(d)} \ln(1 + \frac{c t}{d})}.
\end{equation}
This is precisely the same form as Eq. (\ref{log}). The form of Eq. (\ref{final}) perfectly agrees with the simulations as shown in Fig. 2.

\section{Discussion}
 
We have shown that the discrete lattice model Eq. (\ref{latticemodel}) exhibits an anomalously slow density relaxation. Although our model is quite simple, it contains  essential features of  granular materials that granular  particle can move only when there is enough void below the particle. Tapping the system weakens this constraint and we have taken into account this property by assuming the velocity as Eq. (\ref{vel}). In this way, the logarithmic time--dependence has been obtained, which is consistent with experiments. To our knowledge, no model equation in terms of space--time coordinates, which produces a logarithmic relaxation, has been available so far in granular systems.

The analytical study indicates that the logarithmic relaxation originates from the factor $\exp[-\rho(z)/(1-\rho(z-1)]$ in the velocity Eq. (\ref{vel}). When $\rho(z) \approx \rho(z-1) \approx 1$, the velocity becomes extremely small whereas it is finite when $\rho(z) \approx 0$. Thus we expect that the logarithmic relaxation occurs as far as the system processes this property independent of the detailed form of the velocity. Actually, we have confirmed by simulations that the velocity such as $v = D/(1+e^{(\alpha-1)/\gamma})$ leads to the essentially same relaxation and the existence of a sharp interface in the density profile. This fact implies a universal feature of the logarithmic relaxation not only restricted to granular materials but also for other systems where velocity vanishes abruptly like an essential singularity when the density exceeds some threshold value.

\begin{center}
{\bf ACKNOWLEDGMENTS}
\end{center}

This work was supported by
Grant--in--Aid of Ministry of Education, Science and Culture of Japan. 
G. P. appreciates discussions with Dr. Ruihong Yue and thanks the Japan 
Society for the Promotion of Science.

\begin{figure}
\caption{
Snapshots of density distribution for three different time steps for system with $L = 1024$, $\gamma = 1$ and $D=1$. (a) At $t=0$, the density is randomly distributed with a mean value of $0.50$. (b) At $t=1136$ when the density jump is about 0.8 at the interface between the particle phase and the void phase. (c) At $t=1024,000$ density distribution is very weakly dependent of position in the particle phase. 
}
\label{Fig. 1}

\end{figure}
\begin{figure}
\caption{
Time dependence of the average density. Here $t = 0$ is shifted to the time instant when the calculation of $\rho(t)$ starts. Data points are numerical results from the 1-D lattice model with $L = 1024$, $\gamma = 1$ and $D=1$. Best fit using Eq. (1) is also plotted (curve) with fitting parameters $\rho_{\infty} = 0.989749, \delta \rho_{\infty} = 0.150325, G = 0.230741, \tau = 806.443$. The agreement  is remarkable. 
}
\label{Fig. 2}
\end{figure}
\begin{figure}
\caption{
Density profiles (data points) at two different time steps $t= 10240$ (lower)  (starting from a random initial configuration) and $t = 102400$ (upper) obtained from the 1--D lattice model with $L=1024$, $\gamma = 1.8$, $D=1$. Best fits using Eq. (13) are also shown (curves) with fitting parameters $\rho_1 = 0.960768, s = 5.35083 \times 10^{-4}, \beta = 2.15145$ for $t=10240$ and  $\rho_1 = 0.967761, s = 5.99141 \times 10^{-4}, \beta = 1.8166$ for $t=102400$. 
}
\label{Fig. 3}
\end{figure}

\end{document}